\documentclass{pasj00}

\DeclareRobustCommand{\ion}[2]{\textup{#1\,\textsc{\lowercase{#2}}}}

\graphicspath{{./}}
\newcommand{\sect}[1]{Sect.\,\ref{S:#1}}
\newcommand{\sects}[2]{Sects.\,\ref{S:#1} and \ref{S:#2}}
\newcommand{\fig}[1]{Fig.\,\ref{F:#1}}

\newcommand{\firstfig}[1]{\fig{#1}}
\newcommand{\eqn}[1]{Eqn.\,\ref{E:#1}}

\newcommand{\eqns}[2]{Eqns.\,\ref{E:#1} and \ref{E:#2}}

\newcommand{\graphflex}[4][figure]{\begin{#1}\begin{center}#2\end{center}\caption{#4}\label{F:#3}\end{#1}}
\newcommand{\graph}[3]{\graphwidth[8cm]{#1}{#2}{#3}}
\newcommand{\graphwidthflex}[6][figure]{\graphflex[#1]{#5\includegraphics[width=#4]{#2.eps}}{#3}{#6}}
\newcommand{\graphwidth}[4][9cm]{\graphwidthflex{#2}{#3}{#1}{}{#4}}
\newcommand{\graphfull}[3]{\graphwidth{#1}{#2}{#3}}

\newcommand{\eql}[1]{\begin{equation}#1\end{equation}}
\newcommand{\eqa}[1]{\begin{eqnarray}#1\end{eqnarray}}

\newcommand{\eqi}[1]{$#1$}

\DeclareRobustCommand*{\unit}[1]{\def~{\,}\ensuremath{\mathrm{\,#1}}}
\usepackage{color}
\definecolor{darkgreen}{rgb}{0,0.45,0}

\begin{document}
\SetRunningHead{Ph.-A.~Bourdin, S.~Bingert, H.~Peter}{Coronal loops above an AR --- observation vs. model}
\Received{2014/03/07}
\Accepted{2014/09/30}

\title{Coronal loops above an Active Region --- observation versus model}

\author{%
	Philippe-A. \textsc{Bourdin}\altaffilmark{1,2,3}
	Sven \textsc{Bingert}\altaffilmark{1}
	and
	Hardi \textsc{Peter}\altaffilmark{1}
}
\altaffiltext{1}{Max-Planck-Institut f{\"u}r Sonnensystemforschung, Justus-von-Liebig-Weg 3, 37077 G{\"o}ttingen, Germany}
\email{Bourdin@MPS.mpg.de}
\altaffiltext{2}{Institut f{\"u}r Weltraumforschung, {\"O}sterreichische Akademie der Wissenschaften, Schmiedlstr. 6, 8042 Graz, Austria}
\altaffiltext{3}{Institut f{\"u}r Astrophysik, Universit{\"a}t G{\"o}ttingen, Friedrich-Hund-Platz 1, 37077 G{\"o}ttingen, Germany}

\KeyWords{Sun: corona --- Sun: UV radiation --- Magnetohydrodynamics (MHD) --- Methods: numerical}

\maketitle

\begin{abstract}
We conducted a high-resolution numerical simulation of the solar corona above a stable active region.
The aim is to test the field-line braiding mechanism for a sufficient coronal energy input.
We also check the applicability of scaling laws for coronal loop properties like the temperature and density.
Our 3D-MHD model is driven from below by Hinode observations of the photosphere, in particular a high-cadence time series of line-of-sight magnetograms and horizontal velocities derived from the magnetograms.
This driving applies stress to the magnetic field and thereby delivers magnetic energy into the corona, where currents are induced that heat the coronal plasma by Ohmic dissipation.
We compute synthetic coronal emission that we directly compare to coronal observations of the same active region taken by Hinode.
In the model, coronal loops form at the same places as they are found in coronal observations.
Even the shapes of the synthetic loops in 3D space match those found from a stereoscopic reconstruction based on STEREO spacecraft data.
Some loops turn out to be slightly over-dense in the model, as expected from observations.
This shows that the spatial and temporal distribution of the Ohmic heating produces the structure and dynamics of a coronal loops system close to what is found in observations.
\end{abstract}

\section{Introduction\label{S:core.intro}}

Observations of the solar corona reveal fascinating features, such as thin loops that are bright in extreme ultra violet (EUV) and X-ray wavelengths, see \firstfig{TRACE}.
These structures connect between opposite magnetic polarities that occur for example in an active region (AR).
Current research discusses e.g. what the heating mechanism of these structure is \cite{Klimchuk:2006} and why coronal loops have a roughly constant cross-section \citep{Peter+Bingert:2012}.
These questions can be addressed with large-scale 3D~MHD simulations using high-performance computing.
We are today able to roughly reproduce the 3D structure of EUV-bright coronal loops, as well as the plasma flows within these loops that match to the observed Doppler shifts \citep{Bourdin+al:2013_overview}.
In such a model all necessary coronal plasma quantities are accessible, so that we can deduce synthetic observations and comparable them to real observations delivered by space observatories, like the Hinode satellite \citep{Kosugi+al:2007,Tsuneta+al:2008,Culhane+al:2007}.

\graph{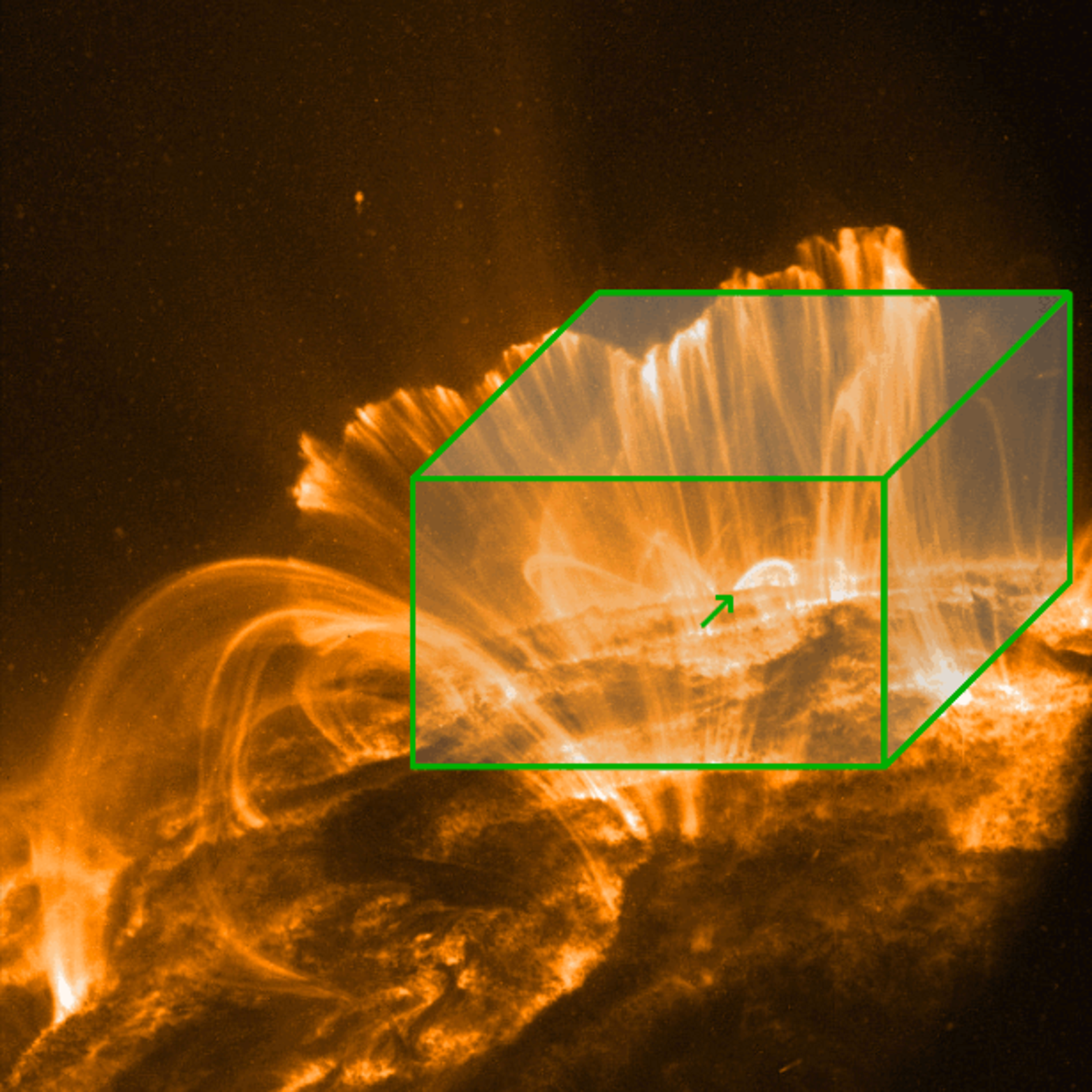}{TRACE}{
Coronal emission in the spectral range of \ion{Fe}{IX} and \ion{Fe}{X}; observed by the {\sc TRACE} space observatory \citep{Strong+al:1994}.
Roughly semi-circular loops span between regions of opposite magnetic polarity in the photosphere.
Hot coronal loops are usually located in the core of an AR and strongly emit in EUV and X-rays; they have typically a smaller and stronger curved structure (see green arrow).
Larger coronal loops are here reaching higher up into the corona; they are usually cooler and hence less bright than the loops in AR cores.
With our 3D model, we want to be able to cover both types of coronal loops and therefore need a sufficiently large computational domain, see overplotted green cuboid.
Image credit: NASA/LMSAL.}

In \fig{TRACE} we display the solar corona with many loops spanning above several ARs and give indications on the size of our 3D~MHD model computational domain.
The goal is to have a large enough box size to capture some of the larger and less bright (cooler) peripheral loops, as well as the smaller loops in the AR core that are brighter and hotter.
We also need to have a resolution in the model (ideally matching the observations) to be able to roughly reproduce also the thickness of such loops that have diameters of about one to several \unit{Mm}.
Previous studies showed a loop-dominated corona in downscaled setups \citep{Gudiksen+Nordlund:2002,Gudiksen+Nordlund:2005a,Gudiksen+Nordlund:2005b}.
In these the energy input was through Ohmic dissipation of currents induced by horizontal motions in the photosphere.
This process was suggested by \cite{Parker:1972} as field-line braiding and later refined by \cite{Priest+al:2002} in their work on flux-tube tectonics.

With scaling laws one can estimate typical parameters of a coronal loop, like the heat input and the plasma pressure.
\cite{Rosner+al:1978} fitted their scaling laws to observed loops, while \cite{Serio+al:1981} added theoretical considerations based on differences between the pressure scaling height and the heat input along a loop.
We can now test such scaling laws by comparing their prediction based on typical loop properties along an ensemble of field lines with the same quantity obtained from the self-consistent 3D MHD model, like the apex temperature or the density.

In observations we find typically smaller hotter loops in the core of an AR, while larger peripheral loops are less intense and cooler (see \fig{TRACE}).
If the heat input in the loops at the photospheric level would be the same, this would contradict the picture of \cite{Rosner+al:1978} and \cite{Serio+al:1981}, where the apex temperature is higher for longer loops for the same heat input.
This lead some authors to assume that larger loops would have the tendency to be hotter \citep{Dowdy+al:1986}.
At least, without a self-consistent description of the spatial dependence of the energy input, the hot cores of ARs would remain elusive.
The density along loops was observed to depend on the loop apex temperature (see \cite{Porter+Klimchuk:1995} and \cite{Aschwanden+al:1999}), which our model confirms.

\section{Coronal model\label{S:model}}

In this study, the box size of 235*235*156\unit{Mm^3} is large enough to encompass a small AR together with some surrounding quiet Sun (QS).
With 1024*1024*256 grid points we achieve a horizontal resolution of 230\unit{km} and a non-equidistant vertical resolution of 100\unit{km} roughly up to the transition region and of 800\unit{km} in the upper corona.
The strong vertical gradients in the temperature and density are resolved well enough to get a substantial match of the model data with co-temporal coronal observations \citep{Bourdin+al:2013_overview}.
In our model we use a constant value for the magnetic resistivity \eqi{\eta} and set its value so that the currents are dissipated on the grid scale (see also \cite{Peter+Bingert:2012}).

\subsection{Boundary conditions\label{S:boundary}}

The lower and upper boundaries are closed for any plasma in- and outflows.
The temperature gradient is forced to be zero, so that no artificial energy transport in or out of the computational domain is possible.
For the temperature, we set the density at the boundary to ensure hydrostatic equilibrium.
In the horizontal direction our model is fully periodic.

\graphfull{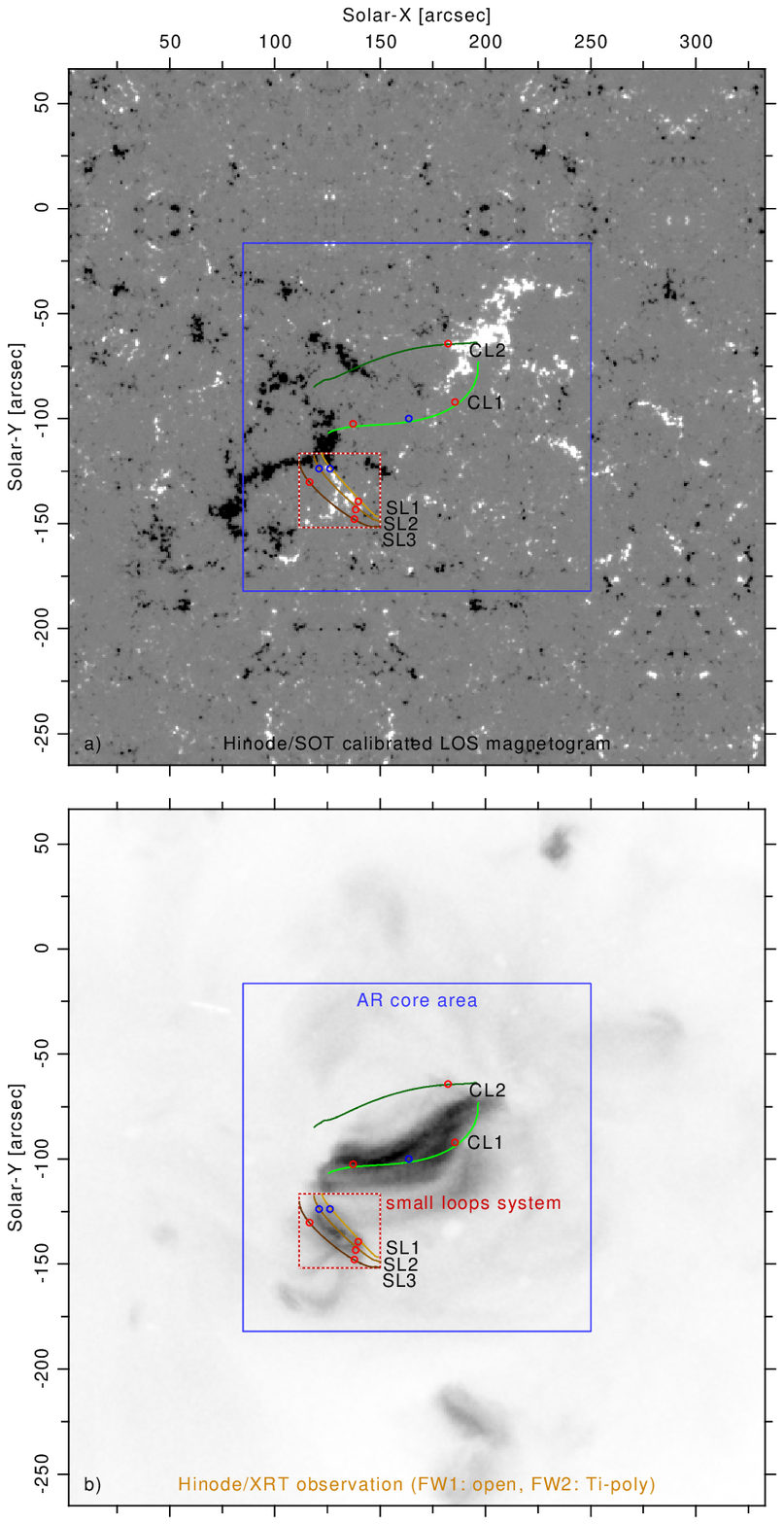}{model.overview}{
Active region observed by Hinode on 14\,Nov\,2007 16:00\unit{UTC}.
In the upper panel we show a photospheric line-of-sight magnetogram taken by SOT/NFI (saturation level: \eqi{\pm 300\unit{G}}) that was calibrated using SP observations performed within the central AR core area (blue quatrate).
The AR magnetogram has a FOV of about 270*160\unit{arcsec^2} and is embedded in the center of a periodic carpet of quiet Sun magnetograms.
The QS magnetograms were calibrated in the same way as the AR with SP data.
Between the different magnetograms we use a 10\unit{arcsec} overlap with a sufficiently smooth transition so that there are no boundary artifacts generated in our simulation results.
The overplotted lines indicate a projection of the magnetic field as computed by the MHD model.
The brightest coronal loops are labeled identical to our previous study (CL\,1--2 and SL\,1--3, see \cite{Bourdin+al:2013_overview}).
In the lower panel we show the coronal X-ray emission observed by XRT.
The circles indicate the footpoint location of the loops, where blue indicates a plasma upflow and red a downflow.}

We prescribe the lower boundary of the model with photospheric magnetograms observed by Hinode/SOT (see upper panel in \firstfig{model.overview}).
We deduce the horizontal velocities of the magnetic patches from the time-series of magnetograms with a local correlation tracker.
This velocity field is used to prescribe the changes of the magnetic field at the bottom layer consistently, but it does not catch the convective motions due to granulation.
Therefore, we need to add a horizontal velocity field that can advect the footpoints of magnetic field lines and inject magnetic energy from the bottom layer.
The motions of the field-line footpoints will produce a vertical Poynting flux that can carry magnetic energy into the corona, where it eventually gets dissipated.
This method was used before to model the field-line braiding mechanism properly \citep{Gudiksen+Nordlund:2002,Bingert+Peter:2011}, where the granular motions match to observed power spectra.
The generated radial outflow from the center of the granules has vorticity that leads to strong flows along the inter-granular lanes.

We use a potential-field extrapolation as lower and upper boundary condition for the magnetic field, so that we will not induce currents due to forcing the magnetic field into a particular state (vertical or closed field boundaries).
This allows also for open magnetic field configurations and the magnetic field vector can have any orientation.

\subsection{Initial condition\label{S:initial}}

The initial condition of the atmosphere consists of a stratified atmosphere close to hydrostatic equilibrium.
An analytical hydrostatic solution is not a stable initial condition, because numerical and analytical derivatives are not identical.
These deviations can ignite tiny compressional waves in the photosphere that gain orders of magnitude in amplitude and result in artificial shock waves when crossing the density gradient towards the corona.

We construct a 1D atmospheric column from partly observed and theoretically derived density and temperature stratifications of the Sun and its atmosphere \citep{Stix:1989,Fontenla+al:1993,November+al:1996}.
We relax this stratified atmosphere in a 1D hydrodynamic simulation using a velocity damping and the same diffusion parameters as in our 3D model.
Then, we transfer this numerically stable solution as initial condition for the 3D model.

The initial condition for the magnetic field consists of a potential-field extrapolation from the observed photospheric magnetogram.
Any perturbations from the driving at the lower boundary need at least the Alfv\'en crossing time to reach the corona and to heat in-situ by Ohmic dissipation of currents.
For that reason we smoothly switch on the radiative losses and the heat conduction at later stages in the simulation.
Not doing so would result in a collapsing corona, because the cooling would be immediate, but the heating is delayed.

\subsection{Switching on\label{S:switching.on}}

\graph{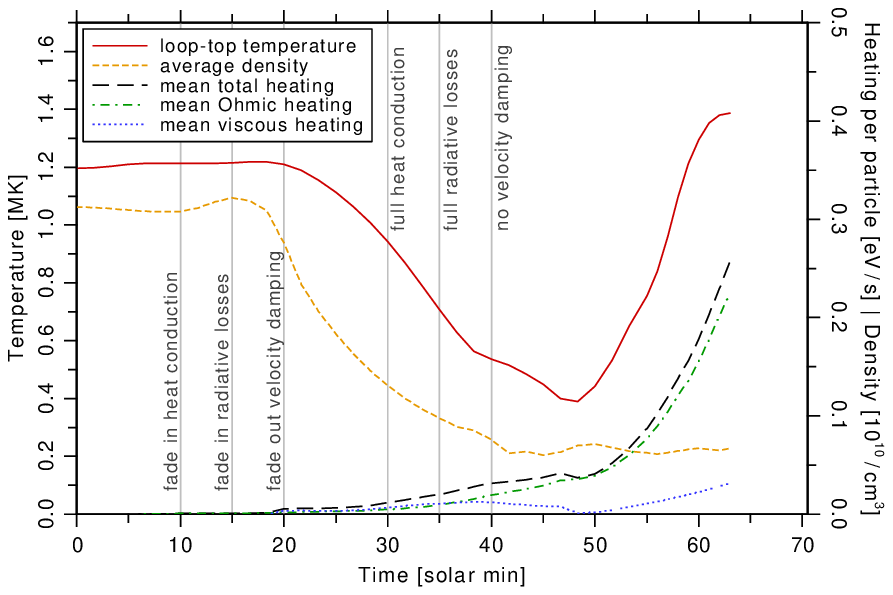}{max_temp}{
Temporal evolution of the maximum temperature and heating terms in the subvolume containing the AR core with the hottest loops (see Fig. 1, CL\,1 and SL\,1 in \cite{Bourdin+al:2013_overview}).
We show here the maximum temperature within the subvolume (red solid line) together with the average Ohmic heating rate per particle (green dash-dotted) and the average viscous heating rate per particle (blue dotted).
The total of both heating terms is shown as black dashed line.
The orange short-dashed line indicates the average density of the coronal part of the hottest loop (SL\,1).
The vertical gray lines indicate points in time when model parameters change, see \sects{initial}{consistent.model}.}

In \firstfig{max_temp} we show how the different terms in the model are switched on.
For example the radiative losses need to be switched on smoothly with a cubic step function, to compensate for the missing part of the Ohmic heating in the coronal energy balance before any magnetic perturbations from the lower boundary have reached the corona.

A strong velocity damping is used to relax any disturbances in the atmospheric stratification that might have arisen from any switch-on effects, like the insertion of the magnetic field and changing model parameters.
The Spitzer-type heat conduction smoothly sets in starting with minute 10 and is fully active at minute 30.
The radiative losses are (in the same way) smoothly switched on from minute 15 to 35.
Then, the whole computational domain starts to cool down due to heat conduction back to the Sun and radiation.
Finally, the velocity damping is smoothly switched off over a time interval of 20\,minutes, spanning from minute 20 to 40, see \fig{max_temp}.

\subsection{Self-consistent model\label{S:consistent.model}}
After 40 minutes, all physical terms are fully active and the velocity damping has faded out, see \fig{max_temp}.
Now the system evolves self-consistently.

After about 48 minutes Ohmic heating starts to become efficient in the corona and the maximum temperature in the subvolume increases rapidly.
This is consistent with the Alfv\'en crossing time through the corona (about 30--40\,minutes) plus several granule life times (5\,minutes each).
Even though the Ohmic heating still rises in the corona, the maximum temperature reaches a kind of plateau.
This is because of two reasons:
First, the heat conduction becomes more efficient with rising temperature, proportional to \eqi{T^{5/2}}, so that it acts like a thermostat, slowing the temperature rise with increasing heat input.
Second, above about 2.5\unit{MK} the optically thin radiative losses also become more efficient with rising temperatures, because of the onset of Bremsstrahlung \citep{Cook+al:1989}.

\subsection{Energy balance\label{S:energy.balance}}
It is important to have a realistic energy balance to self-consistently set the coronal plasma pressure.
For that, we include also coronal energy sinks and transport, like radiative losses \cite{Cook+al:1989} and field-aligned heat conduction \citep{Spitzer:1962}.
The Spitzer-type heat conduction is particularly important to drive chromospheric evaporation at the footpoints of hot loops.

The coronal magnetic field is computed self-consistently within the computational domain.
The equations are written using the vector potential, so that the field always stays divergence-free, even though the vector potential is subject to numeric errors.
To neglect pre-knowledge about the distribution of the current dissipation in the corona, we keep the magnetic diffusivity constant at \eqi{\eta = 10^{10}}\unit{m^2/s}.
With that value we also achieve a Reynolds number around unity for structures on the grid scale, which is required for numerical stability.

\section{Results\label{S:results}}
\subsection{Hot core and cool periphery of the AR\label{S:hot.core.cool.periphery}}

From the model data, we synthesize emission of various emission lines using the atomic database {\sc Chianti} \citep{Dere+al:1997,Young+al:2003} following the procedure from \cite{Peter+al:2004,Peter+al:2006}.
The resulting emission is integrated along the line-of-sight (LOS) because the corona is optically thin.

\graph{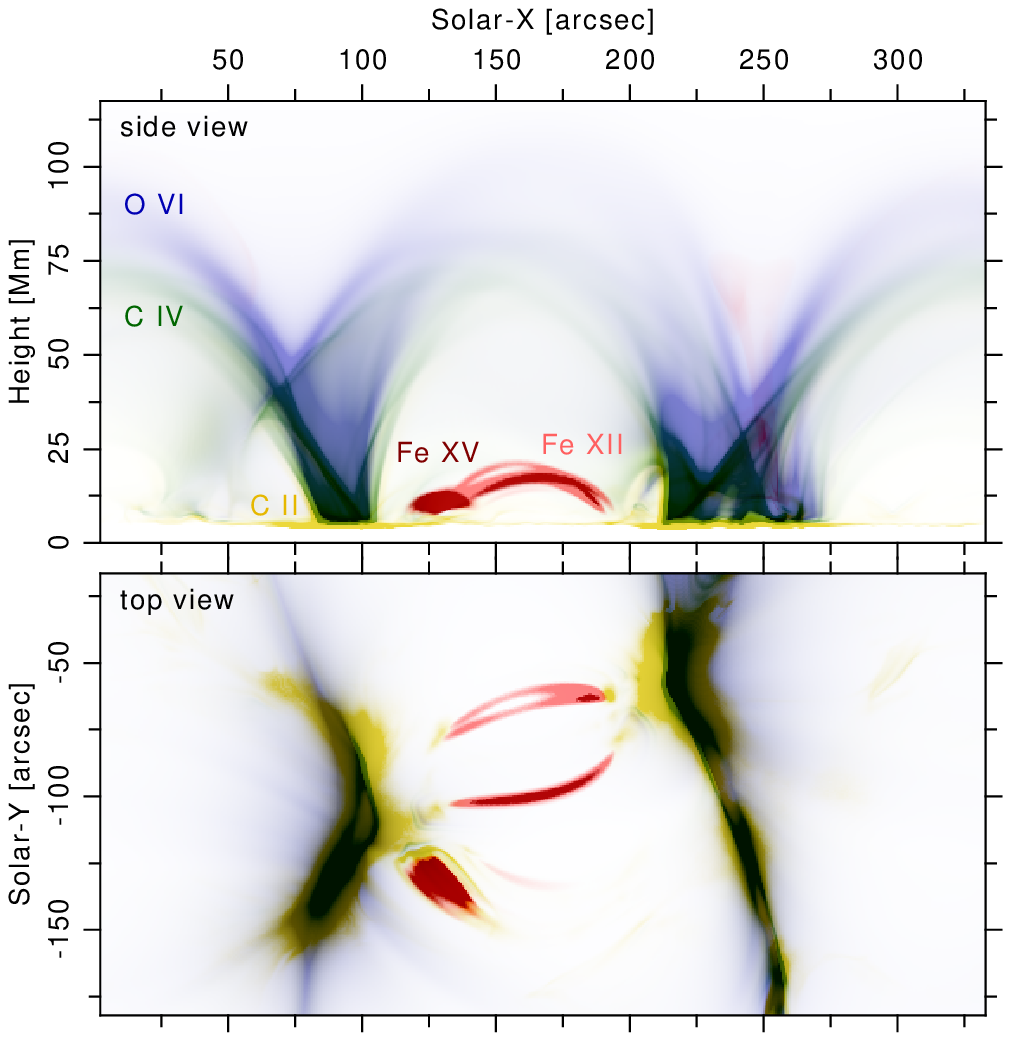}{composite}{
Synthetic emission from the ions \ion{C}{II} (0.04\unit{MK}), \ion{C}{IV} (0.1\unit{MK}), \ion{O}{VI} (0.3\unit{MK}), \ion{Fe}{XII} (1.5\unit{MK}), and \ion{Fe}{XV} (2.5\unit{MK}) as a horizontal LOS integration along the direction of Solar-Y towards north (side view) and as observed at disc center (top view)}.

In \firstfig{composite} we show a composite image of emission in a side-view representing a limb observation of an AR.
We find cooler peripheral loops (blue) that span high over a hot AR core loops system (red).
This trend for cooler loops that span high up into the corona at the periphery of an AR and hot shorter loops within the core of an AR is observed in reality, too, see e.g. the {\sc TRACE} observation shown in \fig{TRACE}.
Our model data qualitatively fits well to this general picture.

\subsection{Scaling laws for coronal loops\label{S:scaling.laws}}

The numerical experiment we perform here can also be used as a test for the scaling laws relating the properties of coronal loops, originally derived from 1D (or even simpler) loop models and observations.
\cite{Rosner+al:1978} derived a scaling-law relation (known as RTV) between general loop properties like the loop length, its maximum (apex) temperature, and the heating deployed along the loop.
\cite{Serio+al:1981} introduced correction factors for the RTV scaling-law to compensate for effects of different pressure and heating scaling heights.

Using an equation of state, in this case the ideal gas law \eqi{p=2 n_e k_B T}, one can derive scaling laws relating the maximum temperature \eqi{T} and the average electron number density \eqi{n_e} to the volumetric heat input \eqi{H} and the loop length \eqi{L}:
\eqa{T &=& c_T \cdot c_H^{-2/7} \cdot H^{2/7} L^{4/7} \cdot E_T {E_H}^{-2/7} \label{E:RTV_T_H} ,\\
n_e &=& \frac{1}{2 k_B} c_H^{-4/7} c_T^{-1} \cdot H^{4/7} L^{1/7} \cdot {E_T}^{-1} {E_H}^{-4/7} \label{E:RTV_n_H} ,}
with the \cite{Serio+al:1981} correction factors:
\eqa{E_T &=& exp\left(-0.04 \cdot L \left(2/s_H + 1/s_P\right)\right) \label{E:Serio_E_T} ,\\
E_H &=& exp\left(0.5 \cdot L \left(1/s_H - 1/s_P\right)\right) \label{E:Serio_E_H} .}

In the original RTV form, \eqi{E_T} and \eqi{E_H} are both equal to \eqi{1} and the constants are:
\eqa{c_T &=& 1400\unit{K (s^2/kg)^{1/3}} ,\\
c_H &=& 9.8 \cdot 10^3\unit{J/m^3}}
in SI units.

We can now test these scaling laws with our model data, if we take the integral of the Ohmic heating along a field line
\eql{F_H = \int {H_{Ohm}(s) \cdot ds}}
and set it equal to the RTV heating \eqi{H} times the loop length \eqi{L}:
\eql{F = H \cdot L \mathrel{\hat=} F_H}
Then, \eqns{RTV_T_H}{RTV_n_H} transform to:
\eqa{T &=& c_T \cdot c_H^{-2/7} \cdot {F_H}^{2/7} L^{2/7} \cdot E_T {E_H}^{-2/7} \label{E:RTV_T_F} ,\\
n_e &=& \frac{1}{2 k_B} c_H^{-4/7} c_T^{-1} \cdot {F_H}^{4/7} L^{-3/7} \cdot {E_T}^{-1} {E_H}^{-4/7} \label{E:RTV_n_F} .}
For the integral \eqi{F_H} we set in the values of the field-line ensemble extracted from our model data.

\subsection{Loop temperature\label{S:loop.temperature}}

We first investigate how well the scaling laws represent the situation in our numerical model with respect to the peak temperature reached along individual field lines.
For this we select an ensemble of about 67'000 field lines in and around the AR core.
Along each field line we extract plasma parameters like the integrated Ohmic heating per particle and the field-line length.
These quantities we use to compute scaling-law predictions of the maximum temperature and average density along loops.

\graph{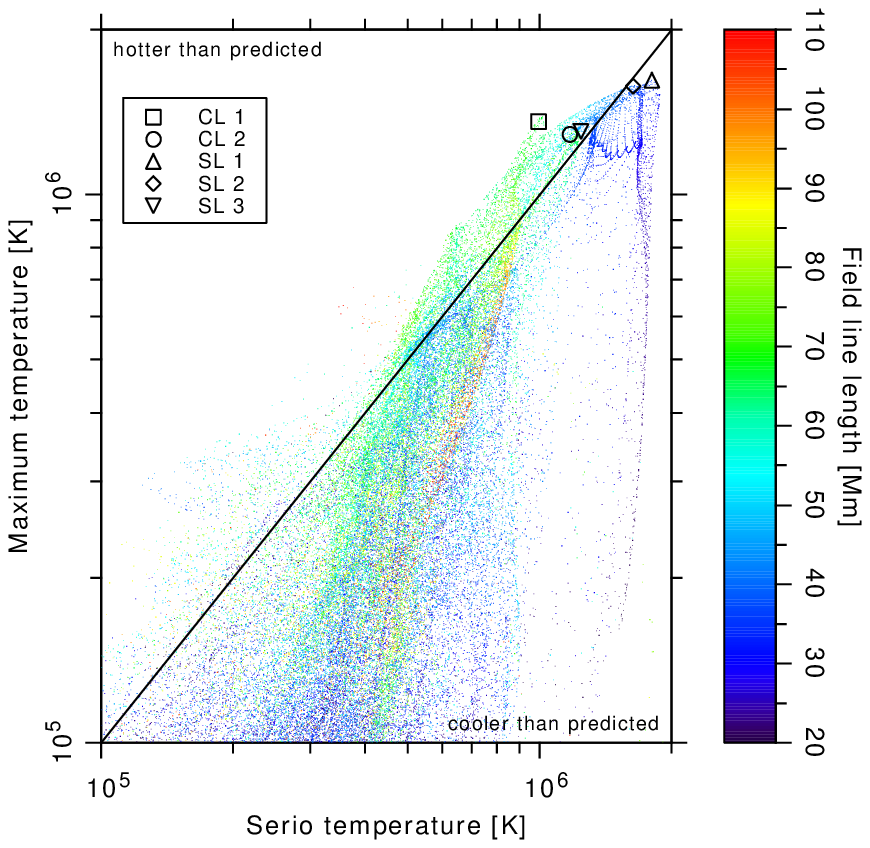}{Temp_Serio_RTV_max_Temp_max}{
Maximum temperature along each field line in an ensemble of 67'000 field lines within our numerical model compared to the respective loop peak temperature according to scaling laws (Serio temperature, see \sect{loop.temperature}).
The black lines indicates were both quantities are equal.}

In \fig{Temp_Serio_RTV_max_Temp_max} we compare the actual peak temperature along the field lines in the sample to the peak temperature as it would be predicted by the scaling law based on the derived heating rate integrated along the field line and the field line length.
We see that the match is reasonable well, even though for low heating rates the temperatures in our numerical model fall short of those predicted by the scaling laws, or in other words, the scaling laws overestimate the loop peak temperatures for small heat input.
This is not surprising, because the loops at these low temperatures (below several 100'000\unit{K}) will not be stable, but intermittent, and are thus not described well by the scaling laws that assume a static loop.

It seems that the agreement between 3D numerical model and observations is particularly good for field lines with lengths of 60 Mm to 80 Mm.
This might (in part) be because these are the loops that could be best observed at the times when the scaling laws were first derived, and thus these scaling laws are sort of best-tuned to loops of such lengths.

\subsection{Loop density\label{S:loop.density}}

\graph{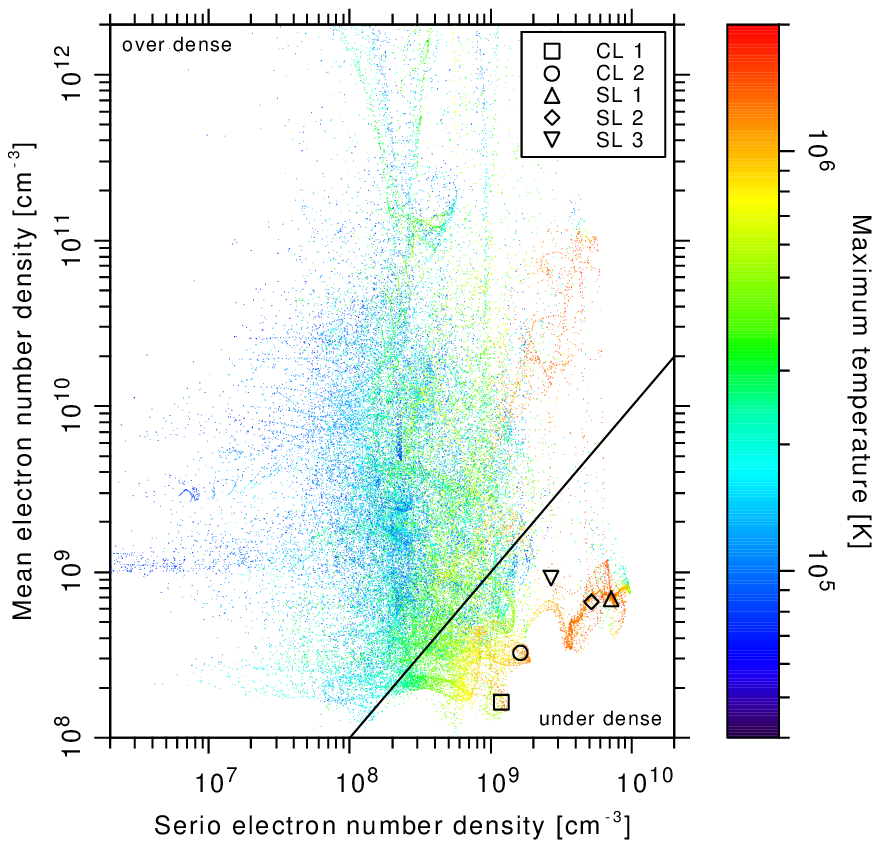}{n_Serio_RTV_rho_n_rho}{
Average (electron number) density of each field line in an ensemble of 67'000 loops within our numerical model compared to the respective loop peak density according to scaling laws (Serio density, see \sect{loop.density}).
The black lines indicates were both quantities are equal.}

We now compare the densities in our numerical model to the densities as they would be predicted for the respective field lines (or loops) based on scaling laws.
The average electron number density in our model field lines is computed as \eqi{<n_e> = L^{-1} \cdot \int n_e(s) \cdot ds}.
In \firstfig{n_Serio_RTV_rho_n_rho} we compare this average number density to the \cite{Serio+al:1981} scaling law number density \eqi{n_e} that is computed from the integrated heating along a field line \eqi{F_{H}}, the correction factors \eqi{E_T} and \eqi{E_H} (\eqns{Serio_E_T}{Serio_E_H}), and the loop length \eqi{L}, see \eqn{RTV_n_F}.
The data points below the equality line (black) are field lines that have a lower density than the scaling law would predict ("under dense"), while the points above are denser than predicted ("over dense") by the scaling laws.

The hottest AR core loops (SL\,1+2 and CL\,1, see \fig{model.overview}) are most under dense, while the warm loops (SL\,3 and CL\,2) are less under dense.
Still, most of our model field lines are over dense, while they are also cooler than predicted, c.f. \fig{Temp_Serio_RTV_max_Temp_max}.
This is consistent with earlier works, where the tendency of cooler loops (\eqi{\sim 1\unit{MK}}) to be over dense was found \citep{Aschwanden+al:1999}, while hotter loops (\eqi{\sim 2\unit{MK}}) were observed with {\sc Yohkoh} and found to be under dense \citep{Porter+Klimchuk:1995}.
For our model loops we find the same tendency, while there is still a minority of data points that do not fulfill this general trend.
We find that our model densities fit best to the values predicted by \cite{Serio+al:1981} around a maximum temperature of roughly 0.8\unit{MK}, which is hence the over/under dense turnover temperature for our model field lines.

The match between numerical experiment and scaling law is much worse for the density than for the temperature (more scatter in \fig{n_Serio_RTV_rho_n_rho} as compared to \fig{Temp_Serio_RTV_max_Temp_max}).
This is clear, because it is much easier to predict the correct temperature for a loop based on a simple (or even a simplistic) model, because the peak temperature depends only weakly on the energy input.
Basically, this is because the heat conduction acts as a thermostat.
The exponent 2/7 with $F_H$ in \eqn{RTV_T_F} stems from $(5/2+1)^{-1}$, where 5/2 is from the Spitzer heat conduction (and $+1$ from an integration).
In contrast, the density (or the pressure) is more sensitive to changes in the heat input, c.f. \eqn{RTV_n_F}.
While the numerical experiment captures this sensitive dependence, the scaling law (which requires averaging) does not, hence the large scatter seen in \fig{n_Serio_RTV_rho_n_rho}.

From our MHD model data we know that the average density is lower inside the hot AR core subvolume than above, which is reflected by the fact that the forming hot loop CL\,1 has an average density lower than most other field lines.
The reason is that the hot AR core subvolume lost some mass towards the chromosphere and was not fed with mass from above, because the magnetic field topology shields this subvolume from incoming plasma downflows.
Altogether, this is perfectly consistent, because if the heating is assumed to be roughly uniform in a coronal loop (in time average), denser loops receive a lower heating per particle, which should actually result in lower temperatures for denser loops as compared to less dense or under dense loops.

\subsection{Coronal energy input\label{S:energy.input}}

Poynting flux is generated by the photospheric driving due to advection of magnetic field by granules and intergranular lanes.
Large-scale motions of magnetic patches and small-scale granular motions introduce magnetic stress at the footpoints of magnetic field lines.
These photospheric perturbations propagate along the magnetic field into the corona and carry energy as Poynting flux.

\graph{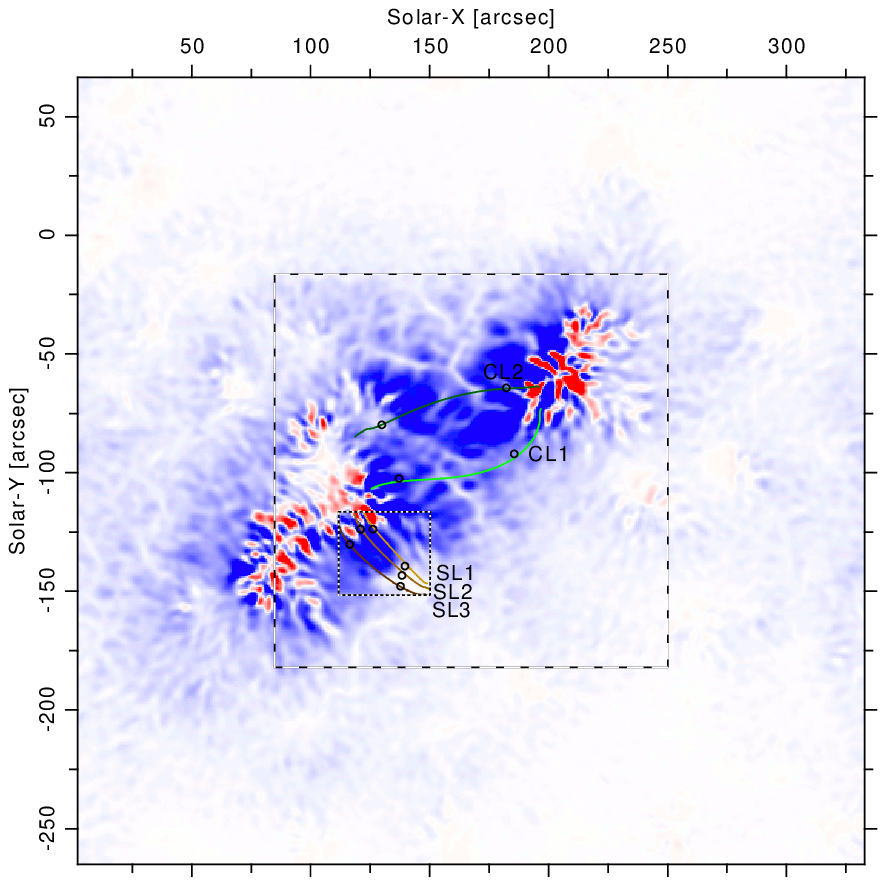}{P_z}{
Vertical component of the Poynting flux at 3\unit{Mm} height.
The footpoints of the AR core loops are marked with circles, the field-line projection is drawn as solid line.
The Poynting flux can be oriented upwards (blue) or downwards (red), the saturation level is set at \eqi{\pm 50'000\unit{W/m^2}}.
The AR core area is indicated as dashed square (c.f. \fig{model.overview}), while the small loops system lies within the dotted rectangle.}

In \firstfig{P_z} we show a horizontal cut through our simulation domain at 3\unit{Mm} height with upwards directed (positive vertical) Poynting flux color coded in blue and downwards (negative vertical) flux in red.
The net vertical Poynting flux not necessarily resembles the real energy input into a field line.
Strictly speaking, the Poynting flux is always perpendicular to the magnetic field.
But, when there is a net positive vertical Poynting flux, an inclined field-aligned structure can be fed from below with magnetic energy.
Therefore, even field lines with a negative vertical net Poynting flux at the legs can have a non-negative energy input, or even the magnetic energy dissipated might come from other sources than the Poynting flux -- such as induction, Lorentz force, or some Poynting flux coming from outside the field line that did not pass through one of the field-line footpoints.

We find in \fig{P_z} that the particularly bright and hot loops CL\,1 and SL\,1+2 are rooted in areas where there is a strong net positive vertical Poynting flux.
Even though some neighboring field lines might have strong negative Poynting flux at their footpoints.
Furthermore, we see that the whole AR core area has a much stronger positive vertical Poynting flux (also after averaging over the neighboring negative flux areas) than the surrounding QS area.
This is clear, because the magnetic flux density is in average much lower in the QS area than in the AR core.

At the base of the corona, the vertical (signed) Poynting flux averages to about \eqi{10^7\unit{erg/s/cm^2}} above the AR core area, see dashed square in \fig{P_z}.

\section{Conclusions\label{S:conclusions}}

From observations we learned that there are hot AR cores with shorter EUV-bright loops, as well as cooler loops spanning high up into the corona that are rooted in the periphery of the AR.
In our 3D MHD model, that has proven to compare well to an observed AR loops system \citep{Bourdin+al:2013_overview}, we also find this trend.

Scaling laws \citep{Rosner+al:1978,Serio+al:1981} predict loop-apex temperatures, as well as the loop pressure (or density).
With our model data, we show that these scaling laws predict well the temperatures of loops with lengths between some 60 to 80\unit{Mm}.
Furthermore, we find that hot loops at temperatures around 2\unit{MK} in our model have a tendency to be under-dense, which was also found in {\sc Yohkoh} observations \citep{Porter+Klimchuk:1995}.
While loops that are cooler than about 1\unit{MK} are typically over-dense, just like observed \citep{Aschwanden+al:1999}.

\cite{Rosner+al:1978,Serio+al:1981} scaling laws would predict that shorter loops are less heated and hence cooler.
In contrast to that, we find a trend for hotter and shorter AR core loops and cooler peripheral loops that was also observed.
Therefore, a higher energy input at the base of the corona is necessary to support the increased heating of hot short loops in an AR core.
In our model, this energy input is provided by a stronger upwards directed Poynting flux within the AR core, as compared to the surrounding QS area.

When we compare the density predicted by the scaling laws, we find strong deviations between model and prediction.
Therefore, a reliable estimate of the coronal loop plasma pressure cannot be made only from scaling laws, which is a prerequisite to predict a realistic coronal energy loss by radiation.
We find that self-consistent 3D models are needed to describe well the coronal energy balance and the formation of EUV-emissive loops in an AR.

Altogether, within the framework of our model we find a good support for the field-line braiding or flux-tube tectonics mechanism \citep{Parker:1972,Priest+al:2002} being the driver for the coronal energy input.
We find in our model that the disturbances in the magnetic field introduced by the horizontal motions in the photosphere are on time scales too long to excite Alfv\'en (AC) waves and instead lead to a quasi-static state change in the coronal magnetic field, leading to a continuous magnetic diffusion and dissipation of magnetic energy on spacial scales of down to 230\unit{km}.
Therefore, our model supports the Ohmic (DC) dissipation of induced currents as coronal heating mechanism above the AR.


\par\addvspace{6pt}
\noindent\small{Acknowledgements.\hskip.5em\itshape{
This work was supported by the International Max-Planck Research School (IMPRS) on Solar System Physics and was partially funded by the Max-Planck-Princeton Center for Plasma Physics (MPPC).
The results of this research have been achieved using the PRACE Research Infrastructure resource \emph{Curie} based in France at TGCC, as well as \emph{JuRoPA} hosted by the J{\"u}lich Supercomputing Centre in Germany.
Preparatory work has been executed at the Kiepenheuer-Institut f{\"u}r Sonnenphysik in Freiburg, as well as on the bwGRiD facility located at the Universit{\"a}t Freiburg, Germany.
We thank Suguru Kamio for his help finding active region observations.
Hinode is a Japanese mission developed, launched, and operated by ISAS/JAXA, in partnership with NAOJ, NASA, and STFC (UK). Additional operational support is provided by ESA and NSC (Norway).
The STEREO/SECCHI data used here were produced by an international consortium of the Naval Research Laboratory (USA), Lockheed Martin Solar and Astrophysics Lab (USA), NASA Goddard Space Flight Center (USA), Rutherford Appleton Laboratory (UK), University of Birmingham (UK), Max-Planck-Institut f\"ur Sonnensystemforschung (Germany), Centre Spatiale de Li\`ege (Belgium), Institut d'Optique Th\'eorique et Appliqu\'ee (France), and Institut d'Astrophysique Spatiale (France).
}}

\bibliography{Literatur}
\bibliographystyle{aa}

\end{document}